\documentclass[aps,amssymb,amsmath,prl,twocolumn,
showpacs, superscriptaddress]{revtex4}
\usepackage{graphicx}% Include figure files
\usepackage{dcolumn}% Align table columns on decimal point
\usepackage{bm}% bold math
\usepackage{subfigure}

\usepackage{amssymb,amsmath,amsfonts,latexsym,graphicx,verbatim,dsfont}
\usepackage[matrix,frame,arrow]{xy}
\usepackage{epsf}

\begin{document}
% Title & Authors
\title{Algorithmic approach to simulate Hamiltonian dynamics and
an NMR simulation of Quantum State Transfer}
\author{Ashok Ajoy}
\email{ashokaj@mit.edu}
\affiliation{Department of Physics and NMR Research Centre,
Indian Institute of Science, Bangalore, India}
\affiliation{Nuclear Science and Engineering Department, Massachusetts
Institute of Technology, Cambridge, MA, USA}
\author{Rama Koteswara Rao}
\affiliation{Department of Physics and NMR Research Centre,
Indian Institute of Science, Bangalore, India}
\author{Anil Kumar}
\affiliation{Department of Physics and NMR Research Centre,
Indian Institute of Science, Bangalore, India}
\author{Pranaw Rungta}
\affiliation{Indian Institute of Science Education and Research Mohali,
Sector-26 Chandigarh, India}
%\date{\today}

%========================================================================================================================================================================

\begin{abstract}
We propose an iterative algorithm to simulate the
dynamics generated by any $n$-qubit Hamiltonian. The simulation
entails decomposing the unitary time evolution operator $U$
(unitary) into a product of different time-step unitaries.
The algorithm product-decomposes $U$ in a chosen operator basis
by identifying a certain symmetry of $U$ that is intimately related to
the number of gates in the decomposition. We illustrate the algorithm by first
obtaining a polynomial
decomposition in the Pauli basis of the $n$-qubit Quantum State Transfer
unitary by Di Franco {\it et. al.} (Phys. Rev. Lett. \textbf{101}, 230502
(2008)) that transports quantum information from one end of a spin chain to the
other; and then implement it in Nuclear Magnetic Resonance to demonstrate
that the decomposition is experimentally viable and well-scaled. We furthur
experimentally test the resilience
of the state transfer to static errors in the coupling
parameters of the simulated
Hamiltonian.  This is done by decomposing and simulating the corresponding
imperfect unitaries.
\end{abstract}
\maketitle

{\it Introduction}: Feynman~\cite{Feynman82} has stated that it should be
possible to
manipulate the Hamiltonian of one quantum system to simulate the dynamics of
another, exponentially faster than a
classical computer. This serves as one of the main motivations for building a
quantum
information processor (QIP)~\cite{lloyd}. More precisely, simulation is the
desire to mimic the unitary dynamics $U$ of a $n$-qubit Hamiltonian
${\cal H}$; and if the physical resources required for the simulation scale
polynomially with the problem size $n$, then it is said to
be {\it efficient}~\cite{lloyd}.

The Hamiltonian of any QIP is the sum of an intrinsic part
${\cal H}_{\rm{int}}$ and a time dependent part ${\cal
H}_{\rm ext}(t)$~\cite{kha} that can be experimentally controlled such that the
following holds to arbitrary precision~\cite{Kitaev}:
\begin{equation}
U\approx U_{\textrm{sim}} = {\cal T}\exp\left[-i\int_{0}^{\tau}
dt({\cal H}_\textrm{int}+{\cal H}_{\textrm{ext}}(t))\right]\;,
\label{eqn:pulse-seq}
\end{equation}
where $\tau$ is the simulation time and ${\cal T}$ is the Dyson time-ordering
operator. A QIP allows universal
control if it can simulate any unitary
dynamics~\cite{clifford-universal}. However finding the requisite efficient set
of
controls ${\cal H}_{\textrm{ext}}$ is in general a challenge~\cite{test-bed}.

In practice, the simulation is constructed by decomposing $U$ into a
product of unitary evolution operators $U_j$:
$U_{\textrm{sim}}=U_1U_2....U_m$~\cite{Kitaev} for a sequence
of time-steps $\delta_j$ with $\sum_{j=1} ^m\delta_j=\tau$. A
polynomial product decomposition (PPD) of $U$ is a decomposition such
that $m$ scales polynomially with the problem size $n$.  Clearly, a PPD
is necessary for a simulation to be
efficient.  Efficiency also entails that each $U_j$ (or gates) be
implemented so that the amount of physical resources (spatial cost)
and the implementation time (temporal cost) are small -- i.e. the total cost
incurred scales polynomially with $m$. Moreover, inevitable decoherence
processes in the QIP~\cite{viola} could impose further efficiency
constraints.

For certain unitaries a PPD may not exist in principle~\cite{Kitaev}.
In this case,
or when the PPD of $U$ is not known, one resorts to approximate methods. This
entails resolving $\tau$ into a finer
time-steps $\Delta \tau$, and then by assuming $\Delta \tau\approx 0$ one
expands $U$
onto a product decomposition (PD). Consequently, both the total cost (the number
of time steps) and the
precision $U\approx U_{\textrm{sim}}$ become $\Delta \tau$ dependent: if one
increases (decreases) the precision, then the total temporal cost also increases
(decreases). For instance, the use ~\cite{lloyd} of Suzuki-Trotter
expansion~\cite{trotter} limits the precision to ${\cal O}({\Delta
\tau}^{3/2})$~\cite{clifford-universal}. This has motivated numerical
optimization
methods to achieve the desired balance between the precision and
time~\cite{grape,smp}, but their inherent computationally intensive nature
restricts them to small $n$, and they generally {\it cannot} be easily extended
to
arbitrary $n$.

In this paper, we not only propose an algorithm to {\it exactly}
product-decompose any $U$, but perhaps more importantly the
algorithm also allows one to search for an efficient PPD. The algorithm first
studies the
support of $U$ in a basis $\textbf{B}$ by representing
it as a {\it vector} $\overline{U}=[c_1, c_2,\dots c_N]$,  since $U=\sum c_j
B_j$, where $B_j\in \textbf{B}$ and $N=4^n$. A PD,
$U=\prod_{k}^m\exp(i\theta_k B_k)$, is then obtained by {\it
  systematically} using the unitaries $\exp(i\theta_k B_k)$ (gates)
generated by the basis operators $B_k$ to iteratively rotate
$\overline{U} \to [1, 0,\dots, 0]$. The PD is derived for arbitrary $n$
in an inductive manner, by extrapolating the symmetry in the PD for
small $n$. In principle, our algorithm allows any basis at the start:
if a PPD is not obtained for one choice, then a different basis maybe
tried. Note however, that any random search for such bases is bound to
be inefficient since the optimal search algorithm of Grover \cite{Grover96} is
non-polynomial and hence inefficient. Since a PPD is realized when the
PD consists of a polynomial number of gates, this may be interpreted to
mean that $U$ has a large symmetry in the sense that it is the
composition of a ``few'' rotations: the Quantum Fourier Transform
unitary~\cite{simon, cory2001} being a famous example.  

We note that our algorithm also enables one to understand how errors
and noise affect the simulation of the unitary $U$.
The errors manifest themselves by reducing the symmetry of $U$ hence
increasing the size of the PD. This can be used to develop appropriate
techniques to control the errors.

Other algorithms for simulation have been suggested previously.
Ref.~\cite{decomp} for example uses Given's rotations ~\cite{Givens,
Householder} to
product-decompose $U$ in terms of C$^{n-1}$NOT and controlled phase gates. Here,
we have explored the usefulness of the Pauli operator basis~\cite{basis}
for the decomposition. The advantage is that gates from this basis
can be
implemented in certain spin-based QIPs in a time optimal
manner~\cite{kha,kha2}. Moreover,
other optimal control techniques~\cite{grape,smp} can also be
modularized in our algorithm to further reduce the total cost of
simulation. Note, however, that the Pauli basis will
not lead to
a PPD for all $U$,  and for a given QIP, there may exist several efficient
bases. Recently \cite{anil} employed our algorithm to experimentally
simulate the Quantum
No-Hiding theorem~\cite{pati}. 

We explain the essential ingredients in the algorithm, including the role of
symmetries, by explicitly product-decomposing (and simulating) the unitary that
causes {\it
quantum state transfer} (QST) in a linear spin-$1/2$ chain~\cite{chris}. QST
allows the chain to act equivalent to a ``wire'' in a spin-based QIP
architecture~\cite{bose1}. Simulation is motivated
because the spin chains, as required for QST, are normally hard to
manufacture \cite{bose1}. We
first obtain a PPD in the Pauli basis of a remarkable QST protocol by Di
Franco {\it et al}~\cite{franco2} that inherently doesn't require the chain to
be
initialized~\cite{franco2,paola2007}. 
We then experimentally simulate it in a Nuclear Magnetic Resonance (NMR) QIP
~\cite{madi1997,suter1,suter-iterative}. 

We also use the PPD to investigate the
protocol's robustness against certain kind of
errors~\cite{robust-QST,stolze-2012}. This is done by introducing specific
errors
in the protocol which is then simulated by our algorithm.  We
explicitly demonstrate how the error results in a PD that scales {\it
faster} than the PPD of the error-free protocol.

{\it The QST protocol \cite{franco2}}:
Given a $n$-qubit chain and couplings between the qubits
$J_j=2J\sqrt{4j(n-j)}$ and $B_j=2J\sqrt{(2j-1)(2n-2j+1)}$, the Ising
Hamiltonian
${\cal H}_I$ of the chain is given as,
 \begin{equation}
 {\cal H}_{I} =
\sum_{j=1}^{n-1}J_{j}Z_{j}Z_{(j+1)} + \sum_{j=1}^{n}B_{j}X_{j}\:,
\label{eqn:Isingham}
\end{equation}
where $\{\mathds{1},X,Y,Z\}$ denote spin-$1/2$
Pauli matrices, eg. $X=\frac{1}{2}\sigma_X$. Let
$|\psi_1\rangle\cdots|\psi_n\rangle$ be the initial state of the chain, then QST
is achieved by evolving $U_{I}=\exp(-i {\cal H}_I t)$ for $t=\frac{\pi}{4J}$:
\begin{eqnarray}
&&U_{I}|\psi_1\rangle|\psi_2\cdots
\psi_{n-1}\rangle|0\rangle\label{eqn:QST}\\
&=& \frac{1}{\sqrt{2}}\left[|0\rangle|\psi_{n-1}\cdots
\psi_2\rangle|\psi_1\rangle +
i|1\rangle|\psi_{n-1}^{\perp}\cdots
\psi_2^{\perp}\rangle(X|\psi_1\rangle)\right]\nonumber
\end{eqnarray}
where $\langle \psi^{\perp}_i|\psi_i\rangle = 0 \:\forall i$. A state locally equivalent to
$|\psi_1\rangle$ is then recovered at the end of the chain by measuring the first
qubit in the Z-basis.

{\it The algorithm}:  Consider a $1$-qubit $U$ as an example. It can
be
represented as a column vector
$\overline{U}_{B}=[u_{\mathds{1}},u_{X},u_{Y},u_{Z}]^{T}$ in the $4$-dimensional
vector space with basis
$\textbf{B}=\{\mathds{1},X,Y,Z\}$; and where the subscript
$B$ denotes that it supports the hyperplane where the vector lies. The
coefficients $u_j$ can be deduced via the orthogonality of the basis: eg.
$u_X=\textrm{Tr}(X^\dagger U)$. More importantly, Clifford algebra ensures that
the basis forms a {\it
group} up to a phase, which we denote as $G_0\equiv\textbf{B}$. We define the
norm (squared length) of $U$ in space $G_0$ as,
\begin{equation}
\|\overline{U}\|_{G_0}=\textrm{Tr}(U^\dagger U)=\sum_{j\in G_0} |u_j|^2 = 1\;.
\end{equation}
We recursively use the fact that one can express $\overline{U}_{G_0}$ in an
effective two-component form
$\overline{U}_{G_0}=[\overline{U}_{G_1},\overline{U}_{\widetilde{G}_1}]^{T}$ by
choosing two orthogonal subspaces of $G_0$, for instance, $G_1=\{\mathds{1},X\}$
and $\widetilde{G}_1=(G_0-G_1)=\{Y,Z\}$.
$G_1$ is necessarily chosen to be a proper {\it subgroup} of $G_0$, and
$\widetilde{G}_1$ is a coset. This allows us to decompose the norm as,
\begin{equation}
\|\overline{U}\|_{G_0}=\|\overline{U}\|_{G_1}+\|\overline{U}
\|_{\widetilde{G}_1}\;.
\end{equation}

The algorithm obtains the PD in terms of the unitaries (gates) generated by the
basis elements of $\textbf{B}$. To do so, it rotates $\overline{U}_{G_0}$ to
$\overline{U}'_{G_1}$, and since $G_1$ is {\it closed} under the product
operation, the elements of $\widetilde{G}_1$ have no further role to play. The
desired rotations are obtained by first product-decomposing $U$ (for example)
into $U^\prime=U_YU_ZU$, where the gates $U_Z\equiv \exp(i\phi_Z Z)$ and
$U_Y\equiv \exp(i\phi_Y Y)$ are chosen with appropriate angles
$\phi_Z$ and $\phi_Y$ such that $U^\prime$ has no support in
$\widetilde{G}_1$ -- we call such gates {\it orthogonal rotations}. To obtain
the angles (equivalently, the time of evolution), note that the product $U_Z U$
in the vector representation is simply the following linear transformation of
${\overline{U}}$,
\begin{equation}
\textrm{Exp}(i\phi
Z):
\overline{U}\rightarrow(\cos\frac{\phi}{2}\textbf{1}_
{4\times4}+i\sin\frac{\phi}{2} \hat{P}_{Z})\overline{U}\;,
\end{equation}
where $\hat{P}_{Z}$ is a
permutation
matrix such that $\overline{\sigma_ZU}=[u_{Z},-iu_{Y},iu_{X},
u_{\mathds{1}}]^{T} = \hat{P}_{Z}\overline{U}$.
We first choose $\phi_Z$ that maximizes $\Delta_{G_1}$, the {\it increase} in
norm of
$U_ZU$ in ${G}_{1}$:
\begin{eqnarray}
 \Delta_{G_1} &=& \|\textrm{Exp}(i\phi_Z
Z)\overline{U}\|_{{G}_{1}} -
\|\overline{U}\|_{{G}_{1}}
\\
&=&\sin^2\frac{\phi_Z}{2}(-\overline{U}^{\dagger}\hat{P}_{R}\overline{U}) +
\frac{1}{2}\sin\phi_Z(-i\overline{U}^{\dagger}\hat{P}_{Z}\hat{P}_{R}\overline{U}
)\nonumber
\end{eqnarray}
where $\hat{P}_{R}=\rm{diag}(\textbf{1}_{G_1}, -\textbf{1}_{\widetilde{G}_1})$
is a reflection matrix about $G_1$. The optimal $\phi_Z$ is then a function of
$U$ \cite{ajoy-supp}:
\begin{equation}
 \theta_{Z}[U]\equiv\phi_Z|_{\rm
max}=\tan^{-1}\left(\frac{-i\overline{U}^{\dagger} \hat
{ P } _ { Z } \hat { P } _ { R }
\overline{U}}{\overline{U}^{\dagger}\hat{P}_{R}\overline{U}}\right)\;,
\label{eqn:angle}
\end{equation}
since $\theta_{Z}[\exp(i\theta_Z Z)U]=0$. Similarly $\phi_Y=\theta_Y$ when
$\theta_Y[\exp(i\theta_Y Y)\exp(i\theta_Z Z)U]=0$. The orthogonal rotations
$U_YU_Z$ rotate ${\overline{U}}_{G_0}$ to
${\overline{U}}^\prime_{G_1}=[\overline{U}^\prime_{G_1},
\textbf{0}_{\widetilde{G}_1}]^T$. The PD $U=U_X U_Y U_Z$ is completed by
obtaining $U_X$ such that $U_XU_YU_ZU=\mathds{1} \equiv [1,0,0,0]^T$.

The above can be generalized to a $n$-qubit unitary $U$. Now basis $\textbf{B}$
contains $4^n$ operators, and to within a phase, $G_0\equiv\textbf{B}$ forms a
group under product operation.
In summary, the PD of the $U$ is achieved by iteratively rotating the vector
$\overline {U}_{G_0}$ via orthogonal rotations to smaller and smaller
(subgroup) subspaces until the PD is completed.

The rotations are essentially deduced by diagonalizing $U$ in two-component
form,  analogous to determining the eigenvectors in
``coupled two-level problems'' in quantum mechanics.
More importantly, the order of the rotations are chosen to minimize the number
of terms (gates) in the PD. This is done through {\it dynamic
programming}~\cite{Bellman03}: at each step, the orthogonal rotation
that causes the maximum
transfer of norm $\Delta_G$ to the desired subspace is chosen {\it first}
(or, the $B_j$ that has
the
maximum ``off-diagonal'' contribution to the two-component matrix is
extinguished first). 

In order to minimize the time to obtain a PD for a $U$ of arbitrary
size, we will product decompose $U$ for small sizes, and identify the
structural symmetry in the form of the PD. We will then use the
pattern identified to inductively extended the PD to any size. We now
demonstrate the above by explicitly product-decomposing the QST
unitary.

{\it The PPD of QST unitary}: First consider the $3$-qubit QST $U_{I}$, which
can be represented as the vector $\overline{U}_{G_0}=1/2\sqrt{2}[1, -i, 1, 1, 1,
-i, i,i]^T$ in the space spanned by $G_0$:
\begin{equation}
\{\mathds{1},X_2,X_1X_3,Y_1Y_3,Z_1Z_3,X_1X_2X_3,Y_1X_2Y_3,Z_1X_2Z_3\}.\nonumber
\end{equation}
Divide $G_0$ into two orthogonal subspaces: a subgroup
$G_1=\{\mathds{1},X_2,Z_1Z_3,Z_1X_2Z_3\}$, and its coset space
${\widetilde{G}_1}=G_0-G_1$. The algorithm generates a single orthogonal
rotation $\exp\left(-i\frac{\pi}{2} Y_1X_2Y_3\right)$ in the space
${\widetilde{G}_1}$ that rotates
$\overline{U}_{G_0}$ to $\overline {U}'_{G_1}=1/2[1, -i, 1, 1, 0, 0, 0,
0]^T$. Now a further iteration gives the following PD:
\begin{equation}
U_{I}=\exp\left(-i\frac{\pi}{2}X_2\right)\exp\left(i\frac
{\pi}{2}Z_1X_2Z_3\right)\exp\left(i\frac{\pi}{2}
Y_1X_2Y_3\right)\;.
\nonumber
\end{equation}
Similarly, the PD for the $5$-qubit $U_I$,
\begin{eqnarray}
U_{I}&=&\exp\left(-i\frac{\pi}{2}X_3\right)\exp\left(i\frac
{\pi}{2}Z_2X_3Z_4\right)\exp\left(i\frac{\pi}{2}
Y_2X_3Y_4\right)\nonumber\\
&\times& \exp\left(i\frac
{\pi}{2}Z_1X_2X_3X_4Z_5\right)\exp\left(i\frac{\pi}{2}
Y_1X_2X_3X_4Y_5\right)\;.\nonumber
\end{eqnarray}
The above PD's are seen to exhibit the following pattern: the QST is
mirror symmetric about the center of the chain~\cite{stolze}. For any equal
division of
$G_0$ into $G_1$ and
$\widetilde{G}_1$, there exists a single orthogonal rotation in
$\widetilde{G}_1$ that causes maximum transfer of the norm such that the
resulting unitary has no support in the coset space. This property is
invariant under further iterations. Thus the decomposition of $U_I$ 
scales linearly with $n$. Inductively, the PPD of an $n$-qubit
$U_I$ is given as,
\begin{widetext}
\begin{equation}
U_{I} =
\prod_{k=1}^{\lfloor n/2 \rfloor}
\exp\left(i\frac{\pi}{2}Y_kY_{(n-k+1)}\bigotimes_{j=k+1}^{
n-k } X_ {j}\right)
\exp\left(i\frac{\pi}{2}Z_{k}Z_{(n-k+1)}\bigotimes_{j=k+1}^{n-k
} X_ {j}
\right)\times\left\{ \begin{array}{lr}
\exp\left(-i\frac{\pi}{2}X_{(n+1)/2}\right)\ \ n\: \rm{odd}\\
\textbf{1}\ \ \ \ \ \ \ \ \ \ \ \ \ \ \ \ \ \ \ \ \ \ \ \ \: \ n\:
\rm{even}\end{array}\right.\label{eqn:gen}
\end{equation}
\end{widetext}

\begin{figure}[!]
\centering
\subfigure[]{\includegraphics[scale=0.29]{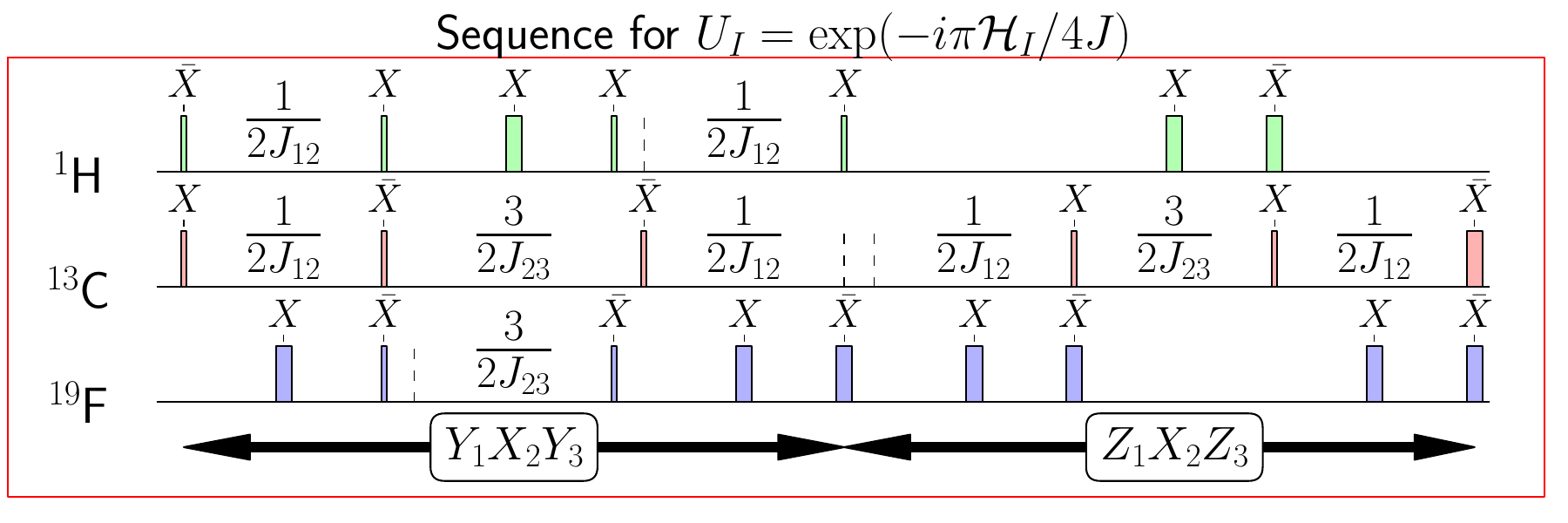}\label{fig:ising2}}
\subfigure[]{\includegraphics[scale=0.29]{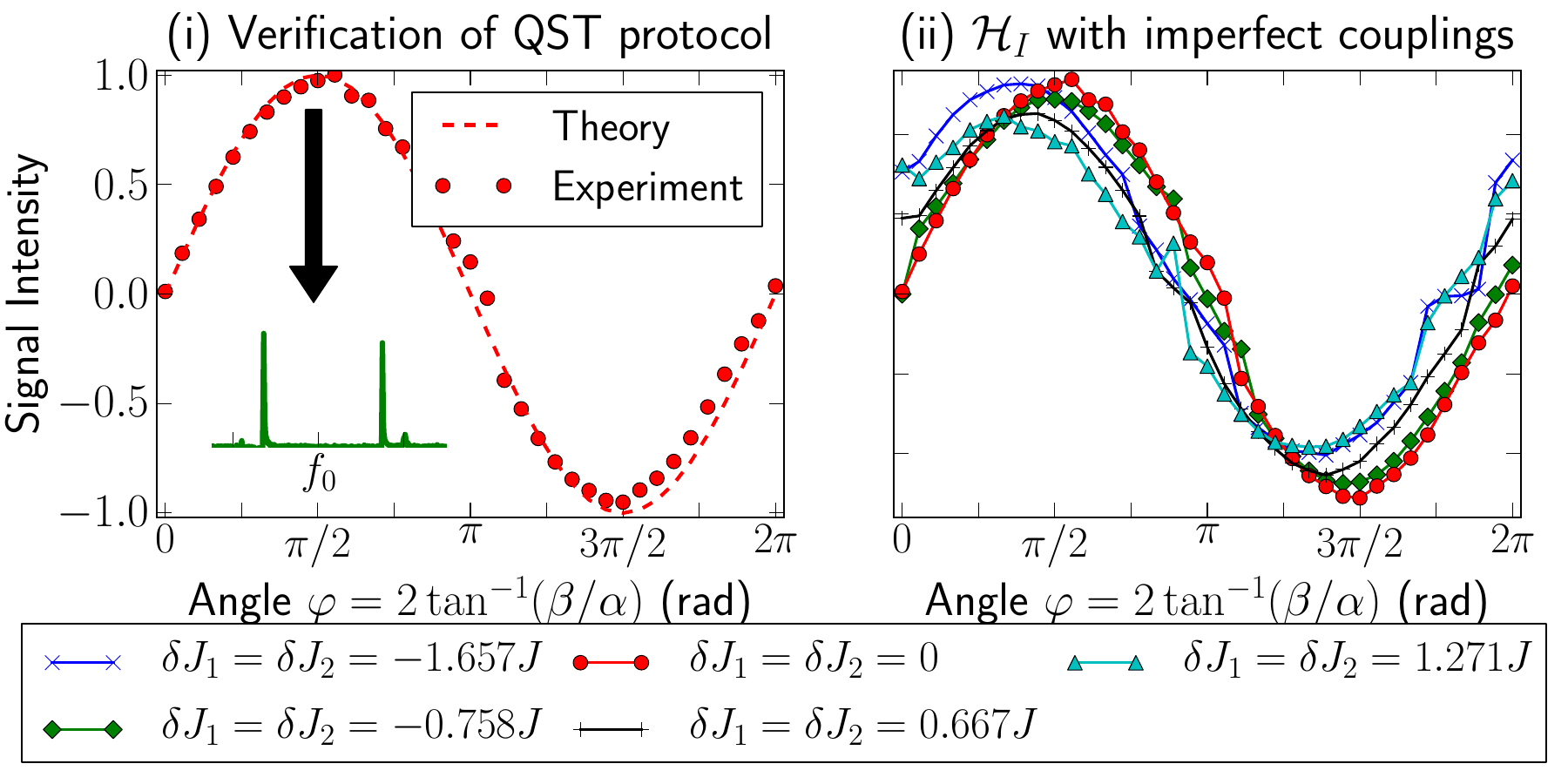}\label{fig:ising3}}
\subfigure[]{\includegraphics[scale=0.29]{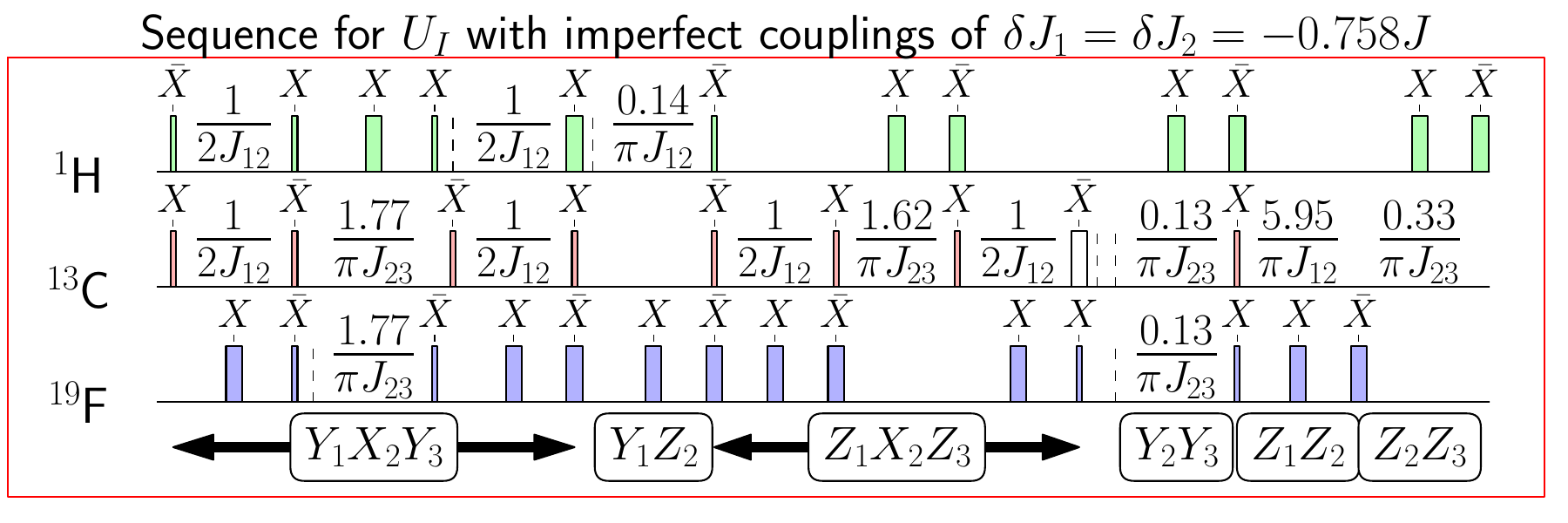}\label{fig:skew1}}
\caption{(Color online) (a) Simplified pulse sequence for $U_{I}$. Filled wide
and
narrow bars are $\pi$ and $\pi/2$ pulses respectively. Dotted lines delineate
periods of Hahn echo
refocusing. Numbers between pulses are delays, with $J_{12}=J_{\rm{HC}}$ and
$J_{23}=|J_{\rm{FC}}|$. (b)(i)
Theoretical (line) and experimental (points) normalized signal intensity on the
third qubit, obtained as the average intensity of the spectral lines
corresponding to the state $(|00\rangle +i|11\rangle)|\psi_I\rangle$ (shown in
the inset for $\varphi=\pi/2$). Here $f_0=-40.26$kHz (-85.6 ppm).
(ii) Experimental signal intensity on the third qubit for different static
errors in ${\cal H}_I$ couplings. A representative pulse sequence
used in (ii) is shown in (c). The unfilled bar is a
103.47$^{\circ}$ $\bar{X}$ pulse.}
\end{figure}

\textit{NMR simulation of $U_I$}:  We used liquid state NMR
to simulate the QST protocol~\cite{suter1,suter-iterative,paola2007} in a
$3$-qubit system
${}^{13}\textrm{CHFBr}_2$, where ${}^{1}$H, ${}^{13}$C, and ${}^{19}$F are
the
qubits. Our motivation is
to study experimental viability of the PD and its scaling.
Experiments were performed at
room temperature in a 11.7~T magnetic field with the resonance
frequencies $500$MHz (${}^{1}$H), $125$MHz (${}^{13}$C), and $470$MHz
(${}^{19}$F). The
couplings between the qubits were $J_{\rm HC}=224.5$Hz, $J_{\rm HF}=49.7$Hz and
$J_{\rm FC}=-310.9$Hz.

The gates in the PPD~(Eq. \ref{eqn:gen}) are implemented by using NMR pulse
sequences~\cite{basis} as the controls. To do so, note that the rotations
generated by any basis operators, such as $B_iB_j$, can be
realized by using ~\cite{clifford-universal}: 
$\exp(-\tau[B_i,B_j]) =
\exp(i\frac{\pi}{2}B_i)\exp(i\tau
B_j)\exp(-i\frac{\pi}{2}B_i)$. Thus orthogonal
rotations by elements of
$\textbf{B}$ are implemented using hard pulses alone, by
Hahn echo refocusing~\cite{ernst1983,tseng}, or by a combination of both.
Piecing together
these sequences leads to the implementation of $U_I$ (Fig. \ref{fig:ising2})
\cite{ajoy-supp}.
In Eq. \ref{eqn:gen} the multiqubit
operators act only on contiguous qubit positions. Hence they can be constructed
only using nearest-neighbor couplings. The non nearest-neighbor couplings
are refocused during the simulation time and
${}^{1}$H, ${}^{13}$C and ${}^{19}$F (in that order) form the requisite
$3$-qubit
chain. 

Let $|\psi_I\rangle = \alpha|0\rangle + \beta|1\rangle$ be the initial state of
the
first qubit, which is to be transferred along the spin chain.
Although the protocol
allows the state of the second qubit to be arbitrary, we fix its initial state to
$|0\rangle$ for ease of measurement. Thus the initial state of the
chain is $|\psi_I00\rangle$,
and is created from the $|000\rangle$ psuedopure state (created by
spatial averaging~\cite{pps}) using a $\varphi=2\tan^{-1}({\beta}/{\alpha})$ X
pulse on the first qubit. The pulse sequence corresponding to the PPD of $U_{I}$
(Fig. \ref{fig:ising2}) is then applied. Since projective measurement is not
possible in an bulk-ensemble QIP like NMR \cite{test-bed}, the readout of the
final state of the
spin chain is done by applying a CNOT$(1,3)$ gate (see \cite{cory-gate}) to
force
the system to $(|00\rangle + i|11\rangle)|\psi_I\rangle$. The state of the third
qubit is measured with the receiver coil in the Y
direction. The results (Fig. \ref{fig:ising3}i) shows excellent agreement with
$\sin\varphi$ \cite{ajoy-supp}, which indeed confirms the QST.

To see the role of symmetry in the study of the robustness of the PPD (thus of
the protocol) to coupling errors, we
consider the simplest case of static disorder in the engineered couplings
\cite{robust-QST,robust-QST2}, where the deviations of the couplings from their
ideal values are independent of qubit position,
\textit{i.e}
$\delta J_1=\delta J_2$. A representative example of the pulse sequence
for this simulation is shown in Fig.~\ref{fig:skew1} \cite{ajoy-supp}. The
experimental results
(Fig.
\ref{fig:ising3}ii) show that the state recovered is corrupted by an additional
phase, and the degree of corruption increases with the error of
couplings. Hence, the transfer fidelity -- characterized by how closely the
final state reproduces the initial state -- decreases with the error of
couplings.

Fig.
\ref{fig:skew1} demonstrates that the PD of the protocol with
coupling error destroys the structural
symmetry~\cite{symmetry} of the error-free PD \cite{ajoy-supp}. Consequently,
the size
of the PD scales faster with the system size, hence incurring higher
simulation cost. In conclusion, our work strongly suggests that the
the structural symmetry of $U$ in a given basis is intimately related
to the scaling of the corresponding PD. Any algorithm that unravels the symmetry
in $U$ can therefore most
economically simulate it.

Authors thank A.D. Patel and P. Cappellaro for
discussions, and K. P. Yogendran for editing the manuscript.
\bibliographystyle{apsrev}
\bibliography{ising}

\end{document}